# Super-resonance effect for high-index sphere immersed in water

Igor V. Minin[a], Song Zhou[b], and Oleg V. Minin[a]
[a]Tomsk Polytechnic University, 30 Lenin Ave., Tomsk 634050 Russia
[b]Jiangsu Key Laboratory of Advanced Manufacturing Technology, Faculty of Mechanical and Material Engineering, Huaiyin Institute of Technology, Huai'an 223003, China.

**Abstract**

*Recently, we showed that dielectric mesoscale spheres support super-resonance effect, i.e. high-order Mie resonance modes with giant field enhancement. The presence of the surrounding medium leads to a significant influence in the intensity of the electric and magnetic fields in the particle. In this paper, we show that this effect can be used for highly precise control of the effective refractive index of a medium, such as water. We show that a change in the water temperature by $\Delta T=0.0106$ °C (or the effective refractive index of the medium by $2*10^{-6}$) leads to a twofold drop in electric field intensity.*

*All the presented results are strictly within the framework of the classical Mie theory without any modifications. A detailed study of the ranges of values of the size parameters of a spherical particle of the order of 10, which had previously been neglected, made it possible to reveal a new, unusual physics of the phenomenon.*

**Introduction**

The role of water in supporting life is extremely important [1,2] and is determined by its optical properties and molecular structure. Water is the key component of plasma of blood, interstitial fluid, intercellular fluid [3], and it is the key and most important liquid of human organism. Knowledge of the optical properties of water and their deviation from the chosen value is very important for solving the problems of biomedical optics. The scattering properties of tissues have a pronounced resonant nature of the wavelength, since the refractive index of the tissue interstitial fluid depends on the refractive index of water [4].

Many different methods for measuring the refractive index (RI) of water have been proposed. The refractive index of water can be determinate by using an empirical formula [5,6] taking into account a knowledge of the wavelength, temperature, and density of the water. RI of water can be measured based on interferometrics and refraction methods [7-11]. Measurement uncertainty is at $10^{-4}$ level. The interferometric methods, based on fringe counting, provide measurement uncertainty about $10^{-5}$. For example, using a frequency comb at 518 nm the refractive index of water with accuracy of measurement uncertainty at $10^{-5}$ level was demonstrated [12].

Recently, a number of new optical phenomena have been discovered for mesoscale dielectric spherical particles with a particle radius on the order of several wavelengths [13,14].

In 2019, we reported that lossless mesoscale spheres, immersed in vacuum, can support high-order internal Mie modes [15-17], which is different from other types of resonances in spherical particles, including WGM [18-22]. These super resonance modes are very sensitive to the Mie size parameter $q$ [15-17] and take place for specific values of this parameter, defined as $q=2\pi R/\lambda$, where $R$ is the radius of particle and $\lambda$ the incident wavelength [23]. Surprisingly, a small energy dissipation in the sphere material can also contribute to sub-diffraction field's localization inside sphere [16] despite of high sensitivity of the super resonance effect to the loss in the particle material.

Recently, it was shown for the first time [24] that the presence of an air around a dielectric sphere, instead of vacuum [15-17], leads to a significant decrease in the magnetic and electric fields intensity enhancement in the particle and a shift in the position of the resonance. Thus, super-resonances modes are extremely sensitive both to the Mie size parameter $q$ and to RI of the environment, and high sensitivity to the loss in the particle material [16]. Below we propose a new physical concept for high precision control of the refractive index of a water, based on the super resonance effect.

**Super-resonance effect for high-index sphere immersed in water**

As a sphere material, we select high index barium titanate glass (BTG) sphere with refractive index of $n_s$=1.90 [25,26] immersed in water, which refractive index depends on several parameters, including temperature [4].

In accordance with the Mie theory [23,27,28], the scattering of a plane, linearly polarized electromagnetic wave on a spherical particle is represented as an infinite series of partial components, each of which can be represented as a sum of two modes - magnetic and electric [27,28].

The coefficients in the equations for electromagnetic field's components are [28]:

$$^l B_n = i^{n+1} \frac{2n+1}{n(n+1)} a_n, \quad ^m B_n = i^{n+1} \frac{2n+1}{n(n+1)} b_n, \quad ^l A_n = i^{n+1} \frac{2n+1}{n(n+1)} c_n, \quad ^m A_n = i^{n+1} \frac{2n+1}{n(n+1)} d_n \quad (1).$$

Electric and magnetic fields inside and outside the particle are expressed through the scattering amplitudes given by $c_n, d_n$ and $a_n, b_n$, respectively (here we are only interested in the internal fields):

$$c_l = \frac{y\zeta_l(x)\psi_l'(x) - y\zeta_l'(x)\psi_l(x)}{y\zeta_l'(x)\psi_l(y) - x\zeta_l(x)\psi_l'(y)} \tag{2}$$

and

$$d_l = \frac{y\zeta_l'(x)\psi_l(x) - y\zeta_l(x)\psi_l'(x)}{y\zeta_l(x)\psi_l'(y) - x\zeta_l'(x)\psi_l(y)}, \tag{3}$$

where $x = k_m a$ and $y = k_p a$,

$$\zeta_l(\rho) = \rho h_l^{(1)}(\rho) = \sqrt{\frac{\pi\rho}{2}} H_{l+\frac{1}{2}}^{(1)}(\rho), \quad \psi_l(\rho) = \rho j_l(\rho) = \sqrt{\frac{\pi\rho}{2}} J_{l+\frac{1}{2}}(\rho),$$

$$\zeta_l'(\rho) = \frac{\partial \zeta_l(\rho)}{\partial \rho}, \quad \psi_l'(\rho) = \frac{\partial \psi_l(\rho)}{\partial \rho}.$$

Here $J_\ell(z)$ and $N_\ell(z)$ are the Bessel and Neumann functions, $k_p, k_m$ - are a wavevectors in medium and particle, respectively. Detailed equations for scattered wave both in medium and inside non-magnetic sphere maybe found, for example, in [28, 29].

In [15-17], the effect of a thousandfold super-enhancement of subwavelength focusing in spherical particles, immersed in vacuum, with certain (more than 10) Mie size parameters and a RI of the sphere less than 2 was discovered. Such spherical particles have a unique arrangement of hot spots at the poles of the sphere, previously mentioned in [20], which in this case are due to the specific behavior of internal Mie modes. With increasing losses in the material of a spherical particle, the maximum intensity enhancement of the electric and magnetic fields are compared [16].

Once super-resonance peaks position and number of modes for selected values of *q* and RI of the medium were found, we simulate based on the Mie theory the magnetic and electric fields intensity distribution in YZ and XZ planes (incident beam x-polarized, propagates from -z to +z direction). The simulation was performed with spatial resolution *a*/200 within both YZ and XZ planes ranging from *-1.2a to 1.2a*, where *a* is radius of particle.

As it was mentioned above, in the Mie theory the behavior of the electric and magnetic fields inside the sphere is fully determined by the partial wave scattering amplitudes $c_l$ (TM mode) and $d_l$ (TE mode) [23,27-32]. The resonances occur at zeros of denominators of $c_l$ and $d_l$ (2,3).

Figure 1 shows the dependence of the Mie the resonant scattering coefficient $\left|{}^m A_{55}\right|$ (1) on the resonant value of the size parameter *q* for mode *l*=55 for several water temperatures (or, equivalently, the refractive index of water). As the temperature rises, the amplitude of the resonance value of the scattering coefficient increases, as does its value (red shifts). The corresponding values of the achievable parameters are shown in Table 1 and Figure 2.

As it follows from Fig.2, the dependence of the resonant value of the size parameter on the water temperature (or on the effective RI of water) is linear. Also one can see from Fig.2 that the dependences of the maximum values of the intensities of the electric and magnetic fields at hot spots vs resonance size parameter are similar to each other, but the intensity of the magnetic field is about 5 times less than that of the electric field.

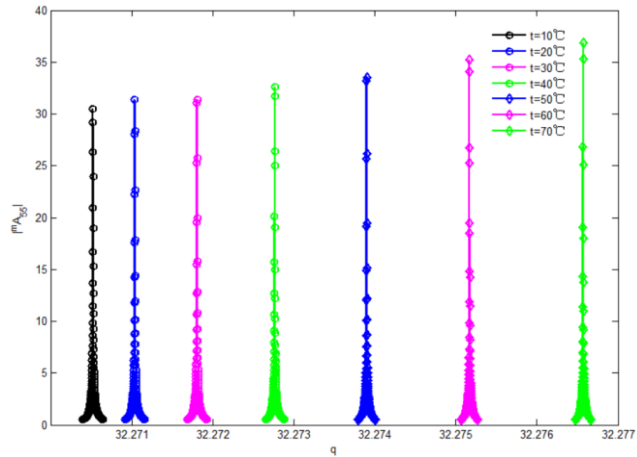

Figure 1. The dependences of the resonant scattering coefficient $\left|{}^m A_{55}\right|$ vs water temperature.

Table 1.

| *t*(°C) | q | $E^2$ (along z-axis) | $H^2$ (along z-axis) |
|---|---|---|---|
| 10 | 32.270524 | 1.817e6 | 3.463e5 |
| 20 | 32.271040 | 1.921e6 | 3.661e5 |
| 30 | 32.271808 | 1.927e6 | 3.669e5 |
| 40 | 32.272766 | 2.073e6 | 3.984e5 |
| 50 | 32.273900 | 2.197e6 | 4.209e5 |
| 60 | 32.275168 | 2.411e6 | 4.672e5 |
| 70 | 32.276570 | 2.642e6 | 5.142e5 |

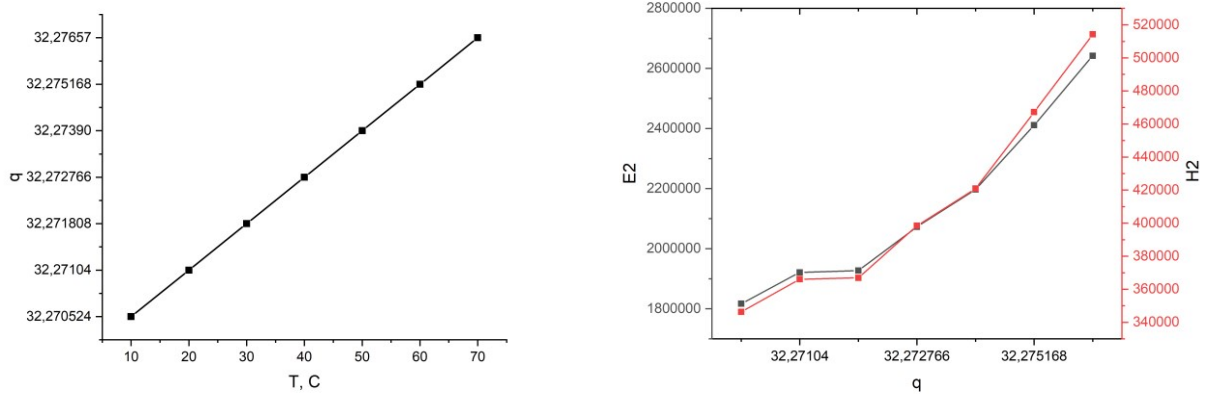

Figure 2. The dependence of the resonant value of the size parameter on the water temperature (left) and the maximum values of the intensities of the electric and magnetic fields at hot spots (right).

An important feature of such mesoscale particles is the possibility of a high degree of field localization, exceeding the diffraction limit, both inside the particle and on its surface [15–17, 24]. This is due to the peculiarity of high-order Fano resonances [13, 15–17] and is associated with the formation of regions with high values of local wave vectors [13], similarly to superoscillation effects [33, 34].

Hot spot configuration for electric and magnetic fields intensity at T=10°C of water temperature are shown in Fig.3. One can see high field localization along the y-direction: simulations shown that for both electric and magnetic fields the FWHM is 0.166λ.

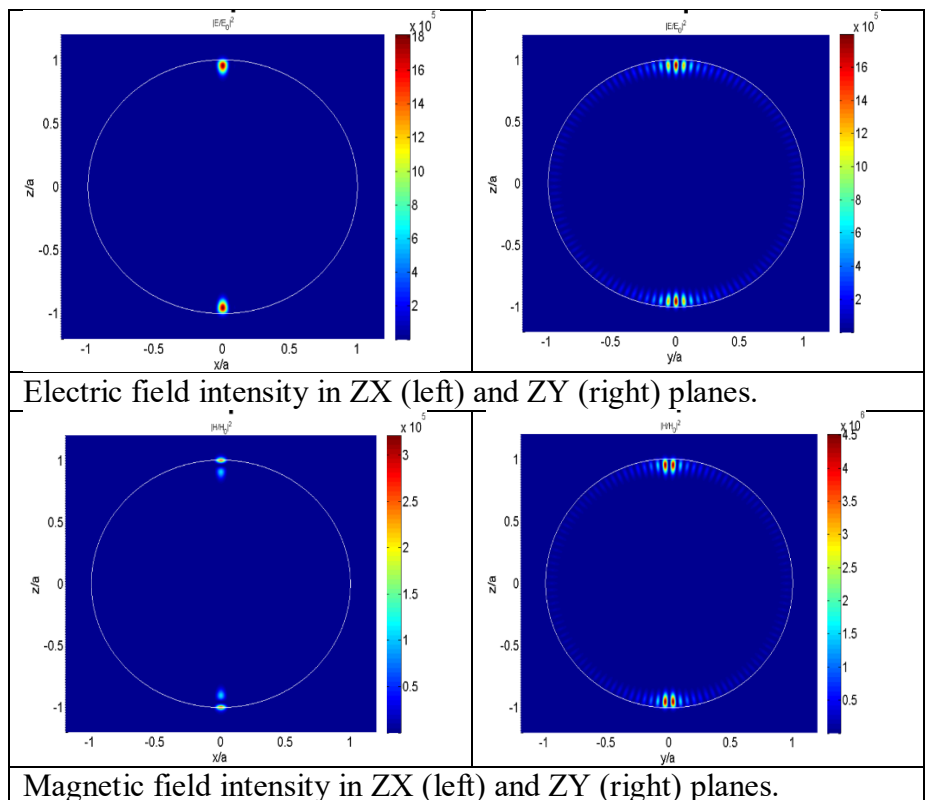

| Electric field intensity in ZX (left) and ZY (right) planes. |
| Magnetic field intensity in ZX (left) and ZY (right) planes. |

Figure 3. Hot spots configuration for electric and magnetic fields at T=10°C of water temperature.

In the case of a dielectric sphere, immersed in water, the high-order internal resonance mode also interferes with a wide spectrum of all other modes. Under certain conditions, a single term with a high-order resonant mode can lead to a multiple increase in the scattered magnetic and electric fields intensity. On Figure 4 this effect is shown for a spherical particle in water and the size parameter q=32.27657. These values correspond to the resonant mode with the number $l$=55.

On Fig. 4a shows the distribution of electric and magnetic fields in two planes, when all modes are taken into account in the simulation up to $l_{max}$=77. On Fig. 4b we present the same results, where all terms are also taken into account, except for the only resonant term with the mode $l$=55. In this case, the electromagnetic fields localization region has a shape characteristic of a photonic jet [35]. Thus, the only term $l$=55 in this case leads to an increase in the intensity of the scattered radiation by a factor of 1000.

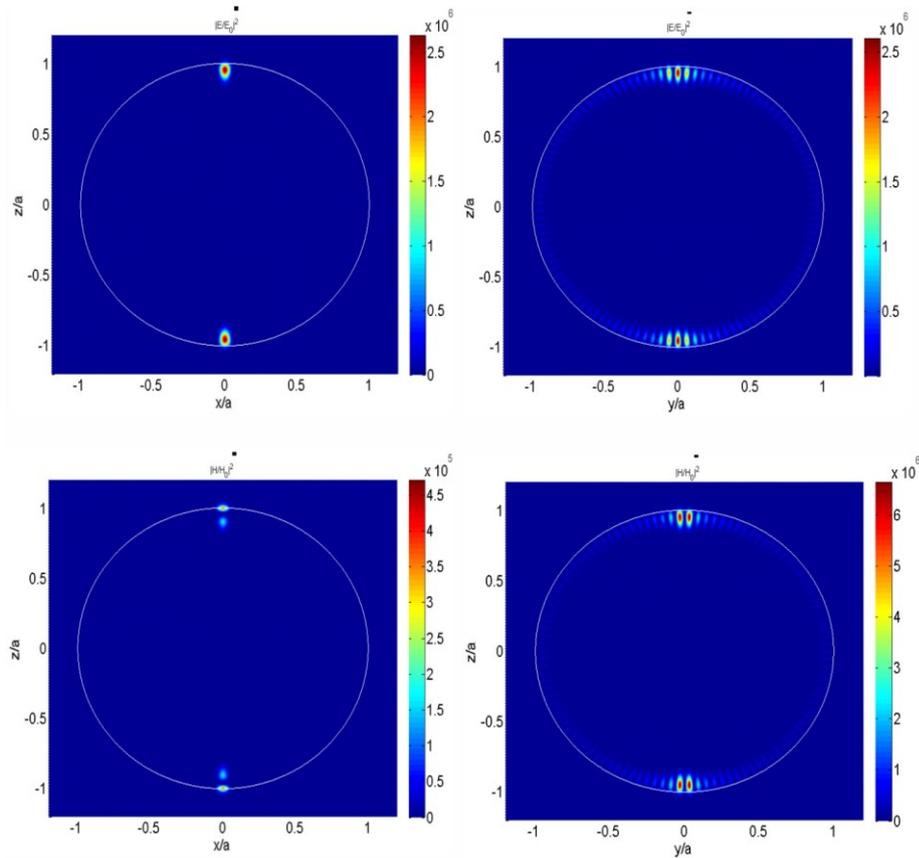

Figure 4a. The distribution of electric (top line) and magnetic (bottom line) fields in two planes, when all modes are taken into account in the simulation up to $l_{max}$=77, for particle immersed in water with T=70°C.

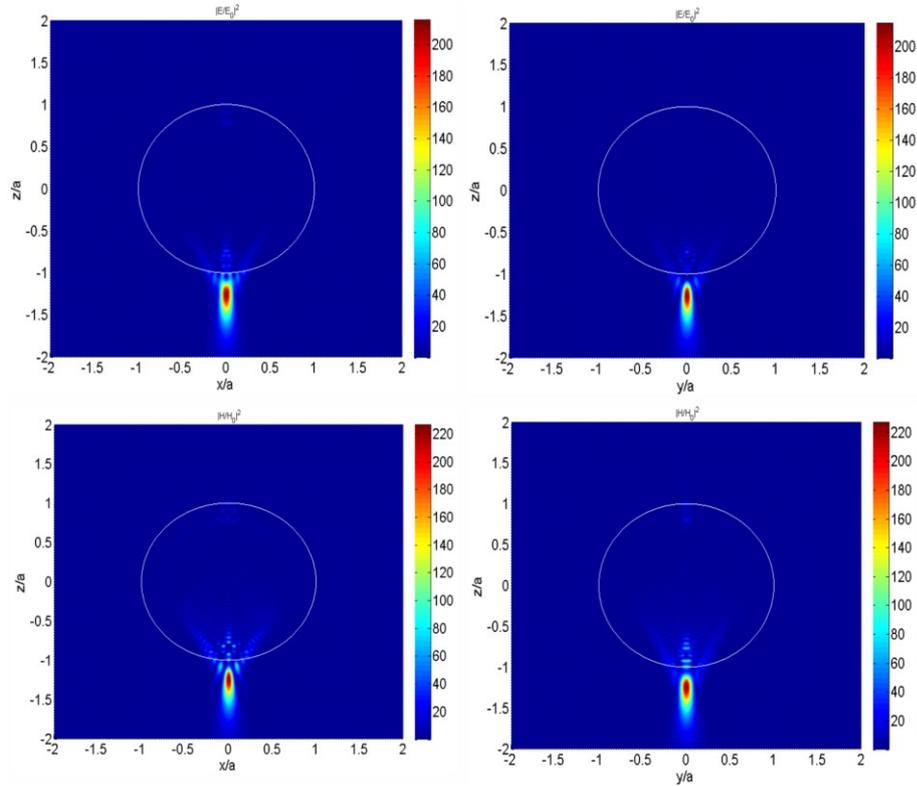

Figure 4b. The same as in Fig.4a except for the only resonant term with the mode $l$=55.

Due to the interference of narrow and broad spectral lines, one can see characteristic Fano line-shapes in the magnetic and electric intensity spectrums on the surface of the particle, i.e. at (x=0; y=0; z=R), as shown in Fig. 4c for sphere with $n_s$=1.9 and $q$~30 immersed in water. The typical range of size parameters necessary to obtain such resonances depend on the refractive index of particle and surrounding medium. On the spectrum of intensities, one can notice the alternation of the prevailing intensities of the magnetic and electric field vs resonant values of $q$. These high order Fano resonances, for which field-intensity enhancement factors in our case can reach values on the order of $10^6$, known as "super resonances" [13-15].

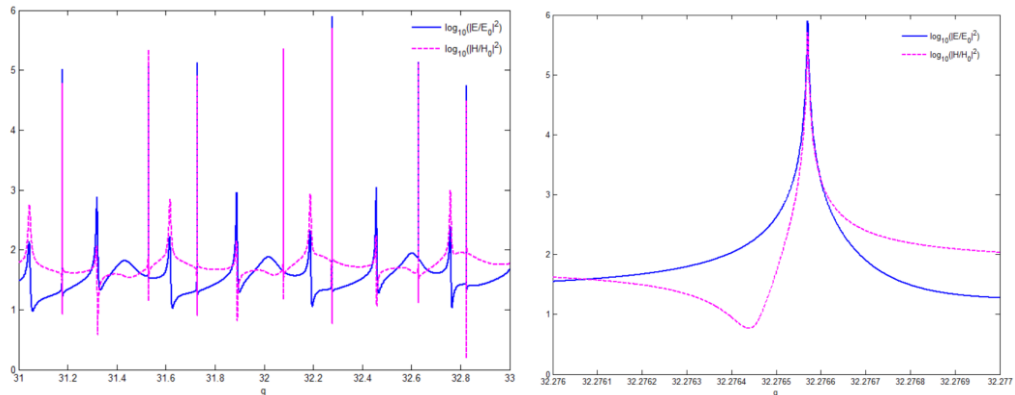

Figure 4c. Magnetic (red) and electric (blue) intensities on the surface for the particle with $n_s$=1.9 and q~30 immersed in water at T=70°C (left). The same near the resonant value of q=32.27657 (right), having the characteristic form of Fano resonance.

As already mentioned above, the super resonance effect is also extremely sensitive to changes in the effective refractive index of the medium [24]. Figure 5 and Table 2 show the hot spot parameters for a small change in the refractive index of water.

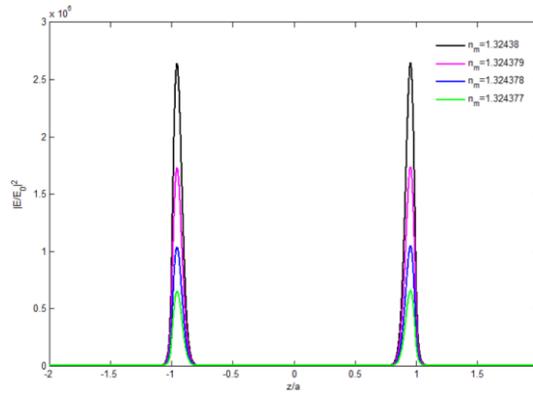

Figure 5. Change in the intensity of the electric field in a hot spot with a small variation in the refractive index of water.

Table 2.

| T(°C) | $n_m$ | $E^2$ (along z-axis) |
|---|---|---|
| 70.0000 | 1.32438 | 2.642e6 |
| 69.9947 | 1.324379 | 1.727e6 |
| 69.9894 | 1.324378 | 1.036e6 |
| 69.9841 | 1.324377 | 6.5e5 |

It is easy to see that a change in the water temperature by ΔT=0.0106 °C (or the effective refractive index of the medium by $2*10^{-6}$) leads to a twofold drop in intensity.

**Conclusion**

In this work, we have investigated a new optical effect [13, 14, 36] - super-resonance modes in mesoscale dielectric spheres, immersed in water, by using Mie theory. Simulations show that a small (about $10^{-6}$) change in the effective water refractive index (for example, which is equivalent to change in water temperature at ΔT=0.01 °C) leads to a twofold decrease in the field intensity at the hot spots of the sphere. This effect can be used for high-precision control of the effective refractive index of liquids, for example, for control of low concentrations of proteins and other biological objects and impurities, in particular, for use in diagnostic systems. It does not matter what reason caused the change in the effective refractive index of water - temperature, pressure, impurities, etc.

The effect of super-resonance for sphere immersed in water may also provide a new route for practical selection of mesoscale particles and surrounding medium to achieve best field sub-diffraction localization and improved focusing resolution, and a new theoretical background to explain the deep super-resolution in real (not virtual) super-resolution images [37] to provide a new operating mode for microsphere-assisted microscopy. We also concluded, using simulation based on Mie theory, that the effective refractive index of the surrounding medium used in such an application is critical to observe the super-resonance effect.

**Acknowledgements**